\documentclass[conference]{IEEEtran}

\usepackage{comment}
\usepackage{url}
\usepackage{algpseudocode}
\usepackage{algorithm}
\usepackage{nth}
\usepackage{amssymb}
\usepackage{epsfig}
\usepackage{threeparttable}
\usepackage{footmisc}
\usepackage{hyperref}
\usepackage{paralist}
\usepackage{wrapfig}
\usepackage{enumitem}
\usepackage{subfig}
\usepackage[table]{xcolor}
\usepackage{color}
\usepackage{multirow}

\usepackage[compact]{titlesec}
\titlespacing{\section}{0pt}{*0}{*0}
\titlespacing{\subsection}{0pt}{*0}{*0}
\titlespacing{\subsubsection}{0pt}{*0}{*0}

\newcommand{\forceindent}{\leavevmode{\parindent=1em\indent}}
\usepackage{graphicx}

\def\BibTeX{{\rm B\kern-.05em{\sc i\kern-.025em b}\kern-.08em
    T\kern-.1667em\lower.7ex\hbox{E}\kern-.125emX}}
\begin{document}

\thispagestyle{plain}
\pagestyle{plain}

\title{Complex Data Analysis Using Multilayer Networks: Modeling,  Efficiency, and Versatility}
\title{Multilayer Networks For Big Data Analytics: Modeling,  Efficiency, and Versatility}
\title{Multilayer Networks For Data-Driven Analysis: Modeling,  Computation, and Versatility}
\title{Multilayer Networks For Data-Driven Analysis: Modeling,  Efficiency, and Versatility}
\title{Data-Driven Analysis: Making a Case for MLNs From Modeling,  Efficiency, and Versatility Perspective}
\title{Why Should We Use MLNs For Data-Driven Analysis? Modeling,  Efficiency, and Versatility}
\title{Making a Case for MLNs for Data-Driven Analysis: Modeling,  Efficiency, and Versatility}

\author{\IEEEauthorblockN{Abhishek Santra\IEEEauthorrefmark{1},
Kanthi Sannappa Komar\IEEEauthorrefmark{2}, Sanjukta Bhowmick\IEEEauthorrefmark{3} and
Sharma Chakravarthy\IEEEauthorrefmark{4}}
\IEEEauthorblockA{\IEEEauthorrefmark{1}\IEEEauthorrefmark{2}\IEEEauthorrefmark{4}IT Lab and CSE Department, University of Texas at Arlington, Arlington, Texas \\
\IEEEauthorrefmark{3}CSE Department, University of North Texas, Denton, Texas \\
Email: \IEEEauthorrefmark{1}abhishek.santra@mavs.uta.edu,
\IEEEauthorrefmark{2}kanthisannappa.komar@mavs.uta.edu,\\
\IEEEauthorrefmark{3}sanjukta.bhowmick@unt.edu,
\IEEEauthorrefmark{4}sharmac@cse.uta.edu}}

\maketitle


\begin{abstract}
\setlength{\parindent}{5ex}
Datasets of real-world applications are characterized by entities of different types, which are defined by multiple features and connected via varied types of relationships. A critical challenge for these datasets is developing models and computations to support {\em flexible analysis}, i.e., the ability to compute varied types of analysis objectives in an {\em efficient} manner.

To address this problem, in this paper, we make a case for modeling such complex data sets as multilayer networks (or MLNs), and argue that MLNs provide a more informative model than the currently popular simple and attribute graphs. Through analyzing communities and hubs on homogeneous and heterogeneous MLNs, we demonstrate the flexibility of the chosen model. We also show that compared to current analysis approaches, a network decoupling-based analysis of MLNs is more efficient and also preserves the structure and result semantics. 


We use three diverse data sets to showcase the effectiveness of modeling them as MLNs and analyzing them using the decoupling-based approach. We use both homogeneous and heterogeneous MLNs for modeling and community and hub computations  for analysis. The data sets are from  US commercial airlines and IMDb, a large international movie data set. Our experimental analysis validate modeling, efficiency of computation, and versatility of the approach. Correctness of results are verified using independently available ground truth. For the data sets used, efficiency improvement is in the range of 64\% to 98\%.
\end{abstract}



\section{Introduction}
\label{sec:introduction}

Real world datasets are composed of diverse types of entities, that are defined by multiple features and interact through varied and complex relationships. With technological advances that allow us to gather increasing amounts of data, "big data" problems are not limited to the size of the data alone, but are also defined by the increasing complexity of the data.

As an example, consider a dataset about a group of actors and directors (these are the entities), each person has some data associated with them, such as who they co-act with, which genre they direct, etc. (these are termed features), actors and directors can also be connected (termed relationships) if an actor is directed by a director. Implicit relations can also be inferred between two entities if they share similar features.


 {\bf Challenges in Multi-Featured Data Analysis.} A critical question is how to efficiently model and analyze  multi-featured datasets that also involve relationships among entities. The move from single feature and/or relationship to multiple features and relationships leads to the following new challenges;

\begin{itemize}
    \item {\em Flexibility of Selecting Features.} Analysis objectives on multi-featured data may require information about a subset of features. For example, given a dataset about  actors, movies in which they acted, and directors who directed the movies (the IMDb dataset used in this paper), the analysis can involve actors and their movie rating, or other actors with whom they work, or movie-rating of actors, or any combination of them. The {\em challenge} is to allow for flexibility of selecting and combining features, while at the  same time avoiding loss of information or redundancy of computations.
    
    
    \item{\em Integrating Analysis of Different Types of Entities.} In addition to variations of the features, the datasets can also contain entities of multiple types. For example, in the IMDb dataset, although both actors and directors are people, they are considered separate types of entities, as they perform different kinds of  jobs. The {\em challenge} is to combine the analysis of these two sets, such as, relating the clusters of actors with some characteristics (co-acting) to clusters of directors with some characteristics (direct-similar-genre).
\end{itemize}


To date, there have not been any generalized framework (a model and computation on the same model) to address these challenges. Most of the work in multi-featured data analysis (see Section~\ref{sec:related-work}) are very focused either on a specific application or a specific analysis technique. There is yet no generalized framework that can capture the myriad necessities of multi-featured data analysis. In view of these challenges, our problem statement is as follows;


{\bf Problem Statement.} \textit{For a given dataset with F features
and T entity types and a set of analysis objectives, develop an framework that (i) generates an expressive model for the data that preserves features and relationships, (ii) allows flexibility of selecting different features, (iii) analysis that includes multiple number of  features and  types of entities, and (iv) enables efficient computation of various analysis objectives.}

{\bf Overview of the Paper.} Our {\em main contribution} in this paper is to present {\em one of the first generalized frameworks for multi-featured, multi-entity datasets}. We present a short survey and comparison of the currently used techniques for modeling and analyzing multi-featured data. We argue that 
 modeling the data using {\em multilayer networks} and then analyzing them using a technique that we term {\em network decoupling} best addresses the challenges discussed earlier. We also demonstrate that by combining these techniques our framework can handle different datasets with multiple features and entity types as well as varied analysis objectives. 

Given below is a brief overview of the approaches to 
 modeling, analysis, mapping and computation, that form our framework.

\setlength{\parindent}{5ex}
    {\em Modeling:} We posit that the model should be able to represent different features of each entity individually or in combination, as well as the relationships between entities. Moreover, the model should also be able to accommodate datasets with similar or dissimilar types of entities.  In Section~\ref{sec:alternatives} we compare the advantages and disadvantages of several standard techniques for modeling of multi-featured, multi-relational data, and show why multilayer networks (MLNs) best fulfill these requirements. 
    
    {\em Analysis:} Many different analysis objectives can be addressed from the same dataset. Computing these objectives requires combinations of different subsets of features or integrating analysis of multiple types of entities.
    In Section~\ref{sec:decoupling} we present {\em network decoupling} method using which information about each feature/entity type is analyzed separately and then the results are combined. Network decoupling provides flexibility of combining layers in multiple ways and also leads to efficient computation.
    
    {\em Mapping:} Modeling and analysis form the backbone of interpreting information from the datasets. However, a critical step is to translate real world objective into analysis expressions.
     For multilayer networks, some of these definitions are more complicated.  In Section~\ref{sec:mapping} we demonstrate with examples how we use analysis objectives to model MLNs and functions for network decoupling.
     
    
    {\em Validation:} In Section~\ref{sec:experiments}, we evaluate our proposed framework, from modeling to analysis to mapping  by validating our results with orthogonal information that were not present in the datasets. We also demonstrate the computational efficiency of our network decoupling approach by comparing the time with the standard network aggregation approach.
\setlength{\parindent}{0ex}


\section{Datasets and Definitions}
We present the datasets we have analyzed and an informal description of some of the relevant terms used for the analysis. 

\subsection{Datasets Used} 

We select two datasets from two different application domains to illustrate the general applicability of our framework. While  much larger datasets  can  be generated, we selected these because reliable ground truth data from orthogonal sources were available. 
The datasets are;



\textbf{1. US-based Airlines: } This is a dataset of six US-based airlines and their flight connections among US cities. This information has been collected by us from multiple sources (\cite{data/Airline/American,data/Airline/Southwest,data/Airline/Delta,data/Airline/Spirit,data/Airline/Frontier,data/Airline/Allegiant}). Here all the entities are of the same type, that is cities. Two cities are related if there is a  direct flight between them. The dataset is characterized by single entity type (city). The multiple features are due to the presence of multiple airlines.

\setlength{\parindent}{5ex}
{\em Analysis Objectives.} We aim to  {\em rank} the top five cities, for each carrier, that have the highest coverage, i.e. can together reach the most number of cities {\bf(A1)};  {\em classify} the airlines into major and minor carriers {\bf(A2)}; and  {\em predict} which city would be selected as its next hub for a carrier planning to expand its coverage {\bf(A3)}.

\setlength{\parindent}{0ex}
\textbf{2. Internet Movie Database (IMDb):} The IMDb dataset is publicly available and stores information about movies, TV episodes, actor, directors, ratings and genres of the movies, etc. \cite{data/type2/IMDb}. Here the entities are of different types as they can be  actors, directors, movies, etc. The features can also differ since actors can be connected if they co-acted or if they work in movies of the same genre.

\setlength{\parindent}{5ex}
{\em Analysis Objectives} We aim to  {\em cluster} groups of co-actors who have worked in movies with high ratings {\bf(A4)};  {\em predict} new groups of actors who have not worked together before, but are likely to work together in future {\bf (A5)};  {\em identify} groups of actors and directors who have close collaborations {\bf (A6)};  {\em refine} the groups found in A6 further to identify actors and directors who have strong collaboration and worked in highly rated movies {\bf (A7)}.

\setlength{\parindent}{0ex}
We selected the analysis objectives to be quite varied. They range from relatively easy analysis of finding coverage of individual airlines and clusters of co-actors to more complicated predictions to the next planned hub of an airline and future potential teaming of actors and directors. 

\subsection{Terminology}
We present some important graph theory concepts  that are relevant to this paper. 



{\em A Network} (or graph), $G$ is an ordered pair $(V, E)$, where $V$ is a set of vertices and $E$ is a set of edges. An edge $(v,u)$ is a 2-element subset of the set $V$. The two vertices that form an edge are said to be neighbors of each other. Here we consider graphs that are undirected (the  vertices in the edge are unordered).

{\em Community Detection} involves identifying groups of vertices that are more connected to each other than to other vertices in the network. This objective is achieved by optimizing network parameters such as modularity ~\cite{Clauset2004} or conductance ~\cite{Leskovec08}. 

{\em Centrality Metrics} are used for measuring the importance of vertices. They include {degree centrality} (number of neighbors), {closeness centrality} (mean distance of the vertex from other vertices), {betweenness centrality} (fraction of shortest paths passing through the vertex), and {eigenvector centrality} (the number of important neighbors of the vertex)~\cite{Newman2010}. In this paper, we use degree and closeness centrality  to quantify the importance of vertices. 

\section{Modeling of Multi-Featured Data}
\label{sec:alternatives}
Multiple relationships among entities or similarity between features of different entities can be concisely expressed as networks. In recent years, network analysis has become a very popular modeling tool for analyzing large datasets of interacting entities. We discuss the different models by which multi-featured data can be expressed as networks, and argue why using multilayer networks would be the best choice.

\subsection{Modeling Complex Data Sets}

{\bf Single Graph or Monoplex:} Here the dataset is represented by a single network or graph. The vertices represent the entities and the edges represent the similarity between the features of the their end points or the dyadic relationships between them.

\setlength{\parindent}{5ex}{\em Advantages.} Modeling data as networks is very  popular because a single feature can be easily expressed as edges in a graph. Moreover, due to extensive research in this area, there exists several algorithms for analyzing them,  such as detecting cliques, communities, hubs, mining subgraphs, motifs etc. Approximations and parallel algorithms for many of these analysis objectives also exist.

 {\em Disadvantages.} Single networks are, however, not adequate in representing multiple features. Particularly, it is difficult to combine features of different categories  (e.g., numerical and categorical), in a meaningful way as one edge. The problem increases when the entities are also of different types.  Moreover, when  analyzing a subset of entities and/or associated feature types, separate graphs may have to be created  for each such combination and analysis.





\noindent{\bf Attribute or Knowledge Graphs:} The simple structure of the networks can be expanded to attribute graphs. Here additional features of the datasets can be represented by including node types in terms of labels (even multiple labels) and multiple edges, even self-loops, corresponding to relationships for different features 

{\em Advantages.} Attribute graphs have been successfully used in
subgraph mining~\cite{tkde/DasC18},
querying~\cite{tkde/JayaramKLYE15},
 and searching~\cite{bigdataconf/HaoC0HBH15} over multi-entity types and multi-featured datasets. By their structure, they capture more semantic information than simple graphs, and can handle both multiple types of features and entities.

{\em Disadvantages.} As  algorithms for some key analysis functions, such as community and hub detection are not yet standardized (or available) for general attribute graphs, they need to be converted to monoplexes for analysis. Although different features can be stored in the graph, for every subset of features, the analysis has to be done separately. This process can lead to redundancy of computations, particularly when the subsets have large overlaps.

 \noindent {\bf Tensors:} The adjacency matrix representation of single networks can be extended to tensors for multilayer networks. A dataset of $V$ entities and $F$ features/relationships, can be represented as a tensor, $A$, of dimensions (V $\times$ F) $\times$ (V $\times$ F). The entry $A^{ia}_{jb}$ gives the connection between vertex $i$ in layer $a$ and vertex $j$ in layer $b$ of the tensor $A$.

 {\em Advantages.} Many network features such as centrality, community using modularity maximization, clustering coefficients can be defined and computed as tensor operations~\cite{Domenico15}. Tensor operations are generally easier than graph based operations to optimize and parallelize for large datasets.

 {\em Disadvantages.} As with the other models, tensors also do not allow for flexible composition of different features, except by analyzing each combination separately. Moreover, tensors are generally used for modeling datasets of one single type of entities and therefore are typically not applied to datasets with multiple types of entities.


\subsection{Modeling as Multilayer Networks}
\setlength{\parindent}{0ex}
Given the pros and cons of these different options, we propose modeling multi-featured, multi-entity type datasets, as multilayer networks (MLNs). Informally, MLNs are layers of single graphs (or monoplexes)\protect\footnote{The terminology used for variants of multilayer networks varies drastically in the literature and many a times is not even consistent with one another. For clarification, please refer to~\cite{MultiLayerSurveyKivelaABGMP13} which provides an excellent comparison of terminology used in the literature, their differences, and usages clearly.}. Each layer, typically, captures the semantics of one particular feature.  As in a monoplex, the vertices of the graph represent the entities of the dataset and the edges represent similarity between the feature values or the dyadic relationship between the end point vertices. The vertices of two layers can also be connected. To differentiate, we term the edges within a layer as intra-layer edges and the edges across the layers as inter-layer edges.

\begin{figure}[h]
\centering
\includegraphics[width=0.9\columnwidth]{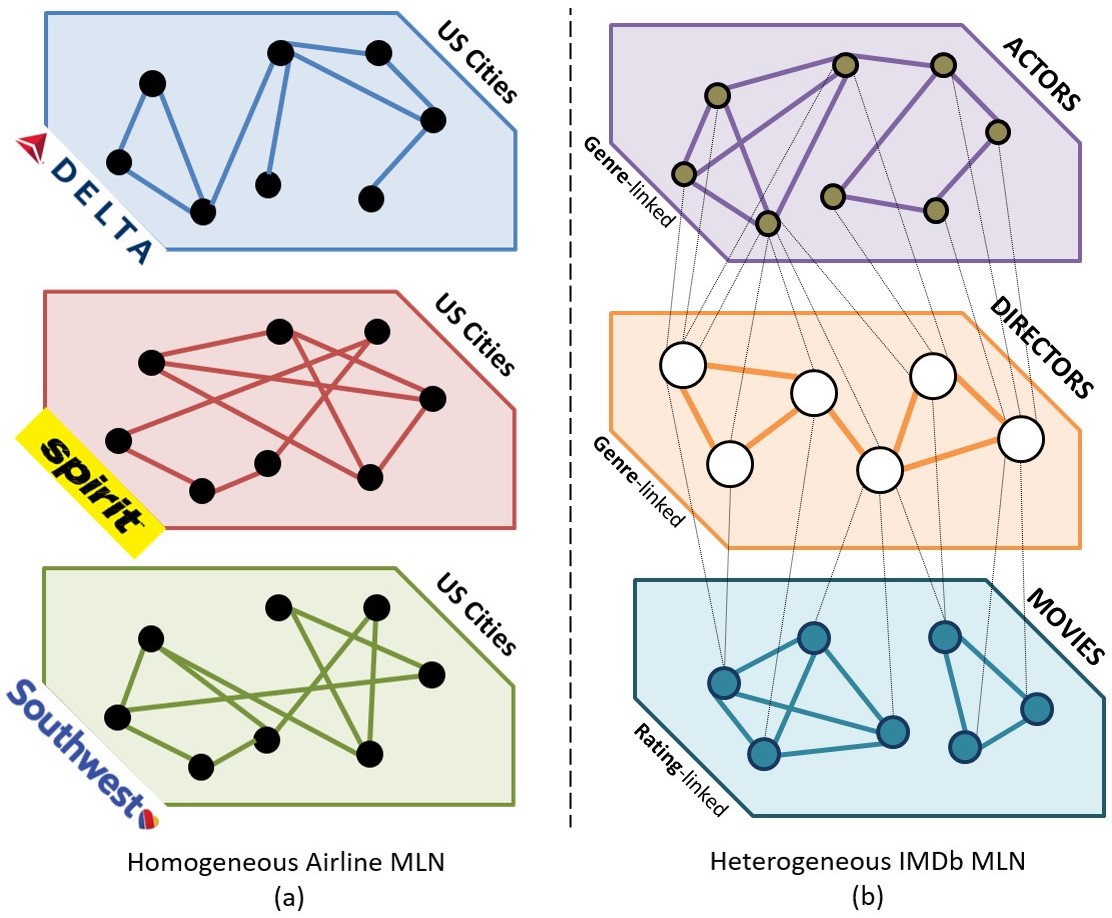}
\caption{Homogenenous and Heterogenenous MLNs}
\vspace{-10pt}
\label{fig:MLN-example}
\end{figure}

    There exists, primarily, two types of multilayer networks -- {\em homogeneous} and {\em heterogeneous}. If each layer of a MLN has the \textbf{same set of entities of the same type}, it is termed a homogeneous MLN (or HoMLN.) For a HoMLN, intra-layer edges are shown explicitly and inter-layer edges are not shown, as they are implicit. US-Airlines is a data set that can be modeled using HoMLN. The nodes in each layer are the same (cities) and each layer can represent a different airline. Within a layer, two nodes (cities) are connected if there is a direct flight between them. Figure~\ref{fig:MLN-example} (a) shows the HoMLN example for the Airline data set.

When the \textbf{set and types of entities are different across layers},then the MLN is termed as a homogeneous multilayer network(HeMLN). The IMDb dataset can be modeled as a heterogeneous multilayer network. Each layer has a different entity type as its nodes (e.g., actors, directors, and movies). The graph of a layer is defined with respect to the chosen features and entity types.

In this case of HeMLNs, the inter-layer links  are defined explicitly based on feature semantics that corresponds to an edge (e.g., directs-actor, directs-movie, acts-in-a-movie). 
Figure~\ref{fig:MLN-example} (b) shows an example from the IMDb data set. 

Further note that whether a data set is modeled as HoMLN or HeMLN depends on the objectives being analyzed. Our choice of IMDb dataset demonstrates this. For example, computation of objectives A4 and A5 can be done using a HoMLN whereas objectives A6 and A7 need a HeMLN.

\setlength{\parindent}{5ex}
{\em Advantages.} Compared to the other options, multilayer networks is a more natural and elegant choice for modeling datasets with multiple entities, features, and relationships.  In MLNs each feature is separately modeled per layer and thus this model can support both heterogeneous and homogeneous datasets.
MLN is also better suited form an information content viewpoint and its visualization.
Incremental changes to each feature or relationship, as modelled by addition/deletion or vertices and edges can be easily included  without extensive re-modeling of the already created MLN.
Unlike most approaches there is no need to convert a MLN representation to another one (simple or attributed) for  analysis when the decoupling approach, discussed in Section~\ref{sec:decoupling}, is used.



{\em Challenges:} Having argued for a MLN for modeling, the primary challenge is to preserve the MLN structure during analysis, thereby preserving both structure and semantics. If this can be done,  further drilling-down of data can be easily accomplished (as shown in Section~\ref{sec:experiments}.). We address this challenge by applying network decoupling.




\section{Analysis of Multilayer Networks (MLNs)}
\label{sec:decoupling}
\setlength{\parindent}{0ex}
We give a brief overview of the current techniques to analyze MLN and discuss how our proposed network decoupling approach improves over the traditional methods.

\subsection{Current Approaches for Analyzing MLNs}
The current approach to analyzing multilayer networks ~\cite{Boccaletti20141,MultiLayerSurveyKivelaABGMP13} is to map the networks to an
 equivalent single graph form. However, through this process, many of the information in the multilayer graphs can be lost.

There are two major techniques for converting a MLN into a single layer network. The first, used for homogenenous networks, is to {\em aggregate the edges of the  multilayer network}. Specifically, given two vertices $v$ and $u$, the edges between them in each layer are aggregated to form a single aggregated edge. This process is repeated for all the vertex pairs.  Some typical aggregation functions are Boolean AND (intersection), OR (union) or linear functions when the edges are weighted. An example, from homogeneous networks, would be aggregating routes of different airplane carriers~\cite{cardillo2013emergence}.




For heterogeneous networks, aggregation is performed in two ways. The first is  {\em type independent}~\cite{LayerAggDomenicoNAL14}, that is ignore the varying types of the entities, and thus basically treat it as a homogenenous network with a subset of vertices in each layer. The second method is  {\em projection-based}~\cite{Berenstein2016, sun2013mining}. Here, if two vertices in a layer are connected to a common vertex in another layer, then an edge can be inferred between them.
Such ``projections" of one layer into another layer, can produce different sets of inferred edges, and then these edge sets are aggregated. An example is connecting drugs that act on common proteins~\cite{Berenstein2016}.

Another method, used for heterogenenous networks, is to {\em transform the multilayer network into an attribute graph}, where the vertices and edges are labeled based on their types. This graph is analyzed to find specified subgraphs, such as patterns of authors, papers and venues~\cite{sun2013mining} or vulnerabilities in infrastructure networks~\cite{Banerjee2016}.

{\em Issues.}
The single network approach has the advantage that many analysis algorithms for community and hub detection are available (e.g., Infomap~\cite{{InfoMap2014}}, Louvain~\cite{DBLP:Louvain} being prominent ones for community detection). However, the aggregation approaches do not preserve either the structure or semantics of MLNs as they aggregate layers. Importantly, aggregation approaches are likely to result in some information loss or distortion of properties~\cite{MultiLayerSurveyKivelaABGMP13} or hide the effect of different entity types and/or different intra- or inter-layer relationship combinations as elaborated in~\cite{DeDomenico201318469}.

In cases, where the multilayer network is converted to an attribute graph,
 the choice of aggregate computations (e.g., community, hub) is limited or may not exist. Some approaches use the multilayer network as a whole~\cite{Wilson:2017:CEM:3122009.3208030} and use inter-layer edges, but do not preserve the layer semantics completely.

\subsection{Proposed Network Decoupling Approach}
Network decoupling is a method by which MLNs can be analyzed without being transformed to the single network form.  The decoupling approach preserves the structure and semantics of the layers as part of the rsult computation, and at the same time can take advantage of the existing algorithms for single layer networks.

\begin{figure}
\centering
\vspace{-10pt}
\includegraphics[width=\columnwidth]{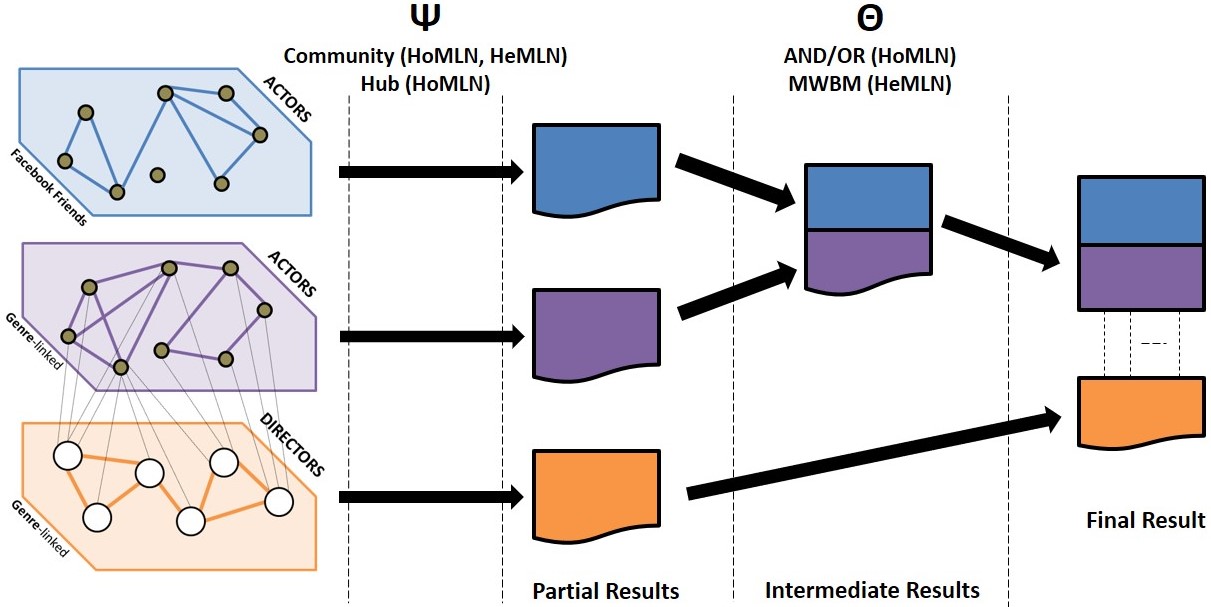}
\caption{Network Decoupling for MLN Analysis}
\vspace{-10pt}
\label{fig:decoupling}
\end{figure}

The {\em network decoupling}  approach as proposed by us in ~\cite{ICCS/SantraBC17,ICDMW/SantraBC17,Arxiv/SantraKBC19} is the equivalent of ``divide and conquer" for MLNs.
This is illustrated in Figure~\ref{fig:decoupling} and is applied as follows, for a given analysis function $\Psi$ and composition function, $\Theta$:
 \begin{itemize}
 \item {\bf(i)} First use the {analysis function $\Psi$} to find analyze each layer of the network individually.

 \item {\bf(ii)} Second, for any two chosen layers, apply a {composition function $\Theta$} to compose the partial results from each layer to generate intermediate or composed partial results.

 \item {\bf(iii)} Finally, apply the composition process until results from all the layers are included.
 \end{itemize}
 Network decoupling has advantages over the traditional aggregation methods.  By using the aggregation approach, information pertaining to the individual layers is lost and it is difficult to measure their relative importance to the system as a whole.  In contrast, network decoupling retains the semantic information of each layer and therefore their individual importance and contribution can be measured.

 The "divide and conquer" approach also facilitates the mix and match of the features and relationships. In the aggregation approach, each time a subset of features is selected, the analysis has to be recomputed, even when the subsets might have overlaps. This leads to redundant computations. Using network decoupling most of the redundant analysis are avoided, since each layer, corresponding to a particular feature is analyzed separately, and then combined.

  \setlength{\parindent}{0ex}
  {\em Challenges.}  Network decoupling has been proposed for both HoMLN~\cite{ICCS/SantraBC17,ICDMW/SantraBC17} and HeMLN~\cite{Arxiv/SantraKBC19}. Moreover, the success of this approach is dependent on correctly matching the analysis function, $\Psi$, with composition function, $\Theta$.  In the next section, using our datasets we will show how we can apply network decoupling to both HoMLN and HeMLN, as well as examples on selecting $\Psi$ and $\Theta$ functions.




\section{Modeling and Analysis of Datasets}
\label{sec:mapping}
In this section we illustrate how we model the test datasets as networks, map the analysis objectives and use network decoupling to perform the computation.

\subsection{Analysis-Driven MLN Modeling}
As indicated earlier, the choice of modeling datasets as HoMLN or HeMLN depends both on the type of entities in the dataset as well as the analysis objectives. If the relationship between entities are not explicitly needed, we formulate and define the similarity between features and thresholds values of similarities based on which edges can be added.  

Below, we show how the identified data sets and analysis objectives discussed in Section~\ref{sec:introduction} are used for  creating MLNs. To recap, the analysis objectives are;
\begin{itemize}
    \item {\em US Airlines}
  \begin{itemize}  
    \item {\bf A1:} rank
the top five cities,
for  each  carrier,  that  have  the  highest  coverage
    \item {\bf A2:} Classify the airlines into major and minor carriers
    \item {\bf A3:} Predict which city would be selected as its next hub for a carrier planning to expand its coverage.
  \end{itemize}  
    \item {\em IMDb Dataset--Actor Collaborations}
     \begin{itemize}  
    \item {\bf A4:} Cluster
groups  of  co-
actors  who  have  worked  in  movies  with  high  ratings

    \item {\bf A5:} predict
new  groups  of  actors  who  have  not  worked  together
before, but are likely to work together in future
\end{itemize}
   \item {\em IMDb Dataset--Actor-Director} Collaborations 
   \begin{itemize} 
    \item {\bf A6:} Identify
groups  of  actors  and  directors  who  have  close  collaborations
\item {\bf A7: } identify actors
and  directors  who  have  strong  collaboration  and  worked  in
highly rated movies
  \end{itemize}
\end{itemize}

\textcolor{blue}{{\em Modeling of US Airline as HoMLN.}} Modeling the US airline dataset is relatively straight forward, since both the nodes and the edges between them are explicitly defined. We model each layer to correspond to a specific airline. We have selected 6 airlines (layers) for analysis -- American, Southwest, Spirit, Delta, Allegiant, and Frontier. Each node in a layer represents a US city. The same set of cities are taken for each airline. Two cities are connected if there is a direct flight between them. In Table~\ref{table:USAirlineHoMLNStats} we give the number of nodes and the number of edges per layer. This MLN is a homogeneous multilayer network, since each layer contains exactly the same set of nodes. 

\begin{table}[h!t]
\renewcommand{\arraystretch}{1}
\centering
    \begin{tabular}{|c|c|c|}
        \hline
        & \textbf{\#Nodes} & \textbf{\#Edges} \\
        \hline
        \textbf{American} & 270 & 746 \\
        \hline
        \textbf{Southwest} & 270 & 717 \\
        \hline
        \textbf{Delta} & 270 & 688 \\
        \hline
        \textbf{Frontier} & 270 & 346 \\
        \hline
        \textbf{Spirit} & 270 & 189 \\
        \hline
        \textbf{Allegient} & 270 & 379 \\
        \hline
        \end{tabular}
\caption{US Airline HoMLN Statistics}
\label{table:USAirlineHoMLNStats}
    
\end{table}


\textcolor{cyan}{{\em Modeling of IMDb as HoMLN.}} For the IMDb network, both types of MLNs will be needed based on different analysis objectives. The analysis objectives {\bf(A4)} and {\bf(A5)} are based on finding clusters of actors, who have either worked together or work in the same genre or work in popular or highly rated movies. Thus the entity set would be the same (set of actors), but the actors would be connected using different features in different layers. Therefore the corresponding MLN would be a homogeneous MLN.

The first layer is the {\em co-acting layer}. Here, two nodes (actors) are connected, if they have {co-acted in at least one movie}. In this layer, the edges are explicitly defined.  

The second layer is the {\em genre layer}. Here two nodes are connected if the actors worked in the same genre. The feature "genre" is a categorical variable, since genres can only take fixed and limited number of values, such as "drama", "action", "comedy", etc. Also note that an actor can act in multiple instances of the same genre -- i.e. in 3 action movies, 1 comedy movie, etc. If we connect actors based on whether they have ever acted in a common genre, the model may not be accurate. For example, a primarily action movie actor, such as Schwarzenegger, could have acted in one or two comedies. 

Inspired by how gene correlation networks are modeled from micro-array data, we evaluate the similarities with respect to genres as follows. For every actor we generate a vector with the number of movies for each genre. We then compute the Pearsons' Correlation Coefficient between the corresponding genre vectors  for each pair of actors. Two actors are connected if the coefficient value is at least 0.9\footnote{The choice of the coefficient  reflects relationship quality. The choice of this value can be based on how actors are weighted against the genres. We have chosen 0.9 for connecting actors in their top genres.}. 

The third layer is the {\em average ratings layer}. This layer models the popularity of the actors based on the ratings of the movies in which they have acted. Two actors are connected if they acted in movies of similar ratings. The movie ratings are given from 0 to 10. Note, however, when we take the average of the ratings, the values become real numbers. To evaluate the similarity we created 10 ranges - [0-1), [1-2), ..., [9-10]. Two actors are connected if their average ratings fall in the same range. Table~\ref{table:IMDbHoMLNStats} gives the details about the three layers.

\begin{table}[h!t]
\renewcommand{\arraystretch}{1}
\centering
    \begin{tabular}{|c|c|c|c|}
        \hline
        & \textbf{Co-Acting} & \textbf{Genre} & \textbf{AvgRating} \\
        \hline
        \textbf{\#Nodes} & 9485 & 9485 & 9485 \\
        \hline
        \textbf{\#Edges} & 45,581 & 996,527 & 13,945,912  \\
        \hline
        \textbf{\#Communities} & 2246 & 63 & 8 \\
        \hline
        \textbf{Avg. Community Size} & 4.2 & 148.5 & 1185.6 \\
        \hline
        \end{tabular}
\caption{IMDB HoMLN Statistics}
\label{table:IMDbHoMLNStats}
    
\end{table}




\textcolor{violet}{{\em Modeling of IMDb as HeMLN.}}  The remaining two analysis objectives on the IMDb dataset, {\bf(A6)} and {\bf(A7)}, relate to connecting directors, actors, and movie ratings. In this case, we need the relationships between these three types of entities and  a separate network layer has to be created for each.  Thus, this network is a heterogeneous MLN.

In the {\em actor layer}, the nodes are actors and two actors are connected if they act in similar genres.  In the {\em director layer}, the nodes are directors and two directors are connected if they direct similar genres. The similarity between genres is computed as in the HoMLN case. 
In the {\em movie layer}, two movies are connected  if they have similar ratings, as per the ranges of the rating values given for the HoMLN network.

Since this is a heterogeneous network, there will also be inter-layer edges. A node from the actor layer is connected to a node from the director layer if the director directed the actor ({\em directs-actor} relation). A node from the actor layer is connected to a node from the movie layer if the actor acted in the movie
({\em acts-in-movie} relation). A node from the movie layer is connected to a node from the director layer if the director directed the movie ({\em directs-movie} relation).  Information about the layers is given in Table~\ref{table:IMDbHeMLNStats}.


\begin{table}[h!t]
\renewcommand{\arraystretch}{1}
\centering
    \begin{tabular}{|c|c|c|c|}
        \hline
        & \textbf{Actor} & \textbf{Director} & \textbf{Movie} \\
        \hline
        \textbf{\#Nodes} & 9485 & 4510 & 7951 \\
        \hline
        \textbf{\#Edges} & 996,527 & 250,845 & 8,777,618  \\
        \hline
        \textbf{\#Communities} & 63 & 61 & 9 \\
        \hline
        \textbf{Avg. Community Size} & 148.5 & 73 & 883.4 \\
        \hline
        \end{tabular}
        \begin{tabular}{|c|c|}
        {\bf Actor-Director Edges} & 32033 \\ \hline
        {\bf Actor-Movie Edges} & 31422 \\ \hline
        {\bf Director-Movie Edges} & 8581 \\ \hline
        \end{tabular}
\caption{IMDB HeMLN Statistics. Top Table: statistics of each layer. Bottom Table: inter-layer edges across layers.}
\label{table:IMDbHeMLNStats}
    
\end{table}

\begin{table}[h]
\centering
        \begin{tabular}{|p{0.6cm}|p{2.8cm}|p{2.8cm}|p{0.9cm}|}
            \hline
            \multirow{2}{0.5cm}{\textbf{Anal-ysis}} & \multicolumn{3}{|c|}{\textbf{Mapping}} \\
            \cline{2-4}
            & \textbf{Computation Order} &  \textbf{$\Psi$} & \textbf{$\Theta$}  \\
            \hline
            \hline
            \multicolumn{4}{|l|}{\textit{US Airline (\textbf{HoMLN})}} \\
            \hline
            \textbf{A1} & Individual layers & Hub (closeness) & none \\
            \hline
            \textbf{A2} & \textit{p} major airline layers; \textit{q} minor airline layers & Hub (degree) & $>$\\
            \hline
            \textbf{A3} & (Target $\Theta$ Competitor) airline layer pair & Hub (closeness) & AND \\
            \hline
            \hline
            \multicolumn{4}{|l|}{\textit{IMDb (\textbf{HoMLN})}} \\    
            \hline
            \textbf{A4} & Co-Acting $\Theta$ AvgRating & Community (Louvain) & AND\\
            \hline
            \textbf{A5} & NOT(Co-Acting) $\Theta$ Genre $\Theta$ AvgRating & Community (Louvain) & AND\\
            \hline
            \hline
            \multicolumn{4}{|l|}{\textit{IMDb (\textbf{HeMLN})} : A (Actors), D (Directors), M (Movies)} \\            
            \hline
            \textbf{A6} & A $\Theta$ D & Community (Louvain) & MWBM\\
            \hline
            \textbf{A7} & \textbf{A} $\Theta$ \textbf{M} $\Theta$ \textbf{D} $\Theta$ \textbf{A} & Community (Louvain) & MWBM\\
            \hline            
\end{tabular}

\caption{MLN Expression for Each Analysis Objective}
\label{table:computation}
\end{table}


\subsection{Network Decoupling on MLNs.}
\label{sec:computation-details}
We describe how network decoupling is applied to the MLNs to compute the objectives. The challenge in successfully applying network decoupling is to match the analysis function, $\Psi$ and the composition function, $\Theta$.

Table \ref{table:computation} gives the mapping of each analysis question {\bf(A1)} to {\bf(A7)} to their actual computation specification (in \textit{left} to \textit{right} order), analysis function ($\Psi$) and composition function ($\Theta$).  Given below is how we determined these functions.

\textcolor{blue}{\em {US Airline Analysis:}} For analysis {\bf(A1)}, our goal is to rank the top five cities for each carrier for coverage. We use closeness centrality to measure the coverage of the airlines, since lower closeness centrality value means lower the average distance between the cities. 

Analysis {\bf(A2)} can be based on the number of connections or more pertinently on the hubs such as degree and closeness centrality. All the three metrics are suitable. We have chosen degree centrality in this case.

Analysis {\bf(A3)} is the most complicated since it requires prediction of future hubs. Here we have to select cities that are not hubs for the airline under consideration and then, compare it with hubs of the other airlines to avoid them. Here too the analysis is done using closeness centrality. We first compute the closeness or degree, individually in each layer. Thus the $\Psi$ function is closeness and degree centrality.

The {\bf second part of the decoupling} process is to combine the results.
{\bf (A1)} computation is with respect to an individual layer and does not require any combination. {\bf(A2)} compares the values across layers. So the composition function $\Theta$ is the greater than ($>$) operation. For {\bf(A3)}, we compare the set of non-hubs, but high closeness centrality vertices, of the carrier to be expanded, with the set of hubs of the competing carriers. Here we can perform an AND operation on the sets to find the common hubs. We discard the common hubs, because they are already the base of the competing carriers.

\textcolor{cyan}{{\em IMDb HoMLN Analysis:}} The analysis objective {\bf (A4)} is to find co-actors who have acted in movies with high ratings. By using network decoupling, we first find communities in the co-actor and  average ratings layer individually. Thus community detection is the analysis function, $\Psi$. We then combine the resultant communities using the composition function AND to obtain groups of actors who have both co-acted together and in high rated movies.

The analysis objective {\bf(A5)} is to find actors who have not acted together but act in the same genre and in movies of similar ratings -- which increases their possibility of acting together in future. We apply the NOT operation on the co-actor layer to find the complement graph of actors who have never acted together. In the first step of network decoupling, we take communities from each of the three layers; the complement of the co-actor layer, the genre layer and the average ratings layer. We then combine the resultant communities using the composition function AND to find groups of actors who have a high chance of acting together in future.  Since we had already computed communities of the average ratings layer as part of {\bf(A4)}, we need not recompute these again. This highlights a benefit of network decoupling.

\textcolor{violet}{{\em  IMDb HeMLN Analysis:}} For the objectives related to this MLN we have to find the communities of actors and directors (objective {\bf(A6)}) and communities of actors, directors and movies (objective {\bf (A7)}) in the individual networks. Thus the analysis function is again community detection. Since we had already computed the communities of actors and directors as part of {\bf(A6)}, we need not recompute them in {\bf (A7)}. 

Finding the connections between the communities is challenging since the entities are of different types. Each community is considered to be a meta-node.  Two meta-nodes in two different layer are connected if there is at least one intra-edge between them. The weight of the edges (meta-edges) between the meta-nodes is given by the number of intra-edges between them. This construction creates a bipartite graph. These meta nodes (communities) in the bipartite graph are paired using the composition function ($\Theta$) {Maximum Weighted Bipartite Matching (MWBM) as proposed by Jack Edmonds}~\cite{edmonds1965maximum}.

{\em Communities in MLNs.} Using the IMDb dataset, we showed two examples of community detection in MLNs. For composition of HoMLNs, with unweighted edges, we can use Boolean AND, OR, and unary NOT. Once we obtain the communities from each layer we combine the communities with the corresponding Boolean operation similar to the algorithm in ~\cite{ICCS/SantraBC17}. 


For HeMLNs, we use a  \textit{structure-preserving HeMLN community detection} from ~\cite{Arxiv/SantraKBC19} that takes into account the {combined effect of layer communities, entity types, intra- and inter-layer relationships (types)}. A community bipartite graph is built with the communities from individual layers, as discussed above, and then matched to reflect the semantics of communities. 

\section{Experimental Analysis}
\label{sec:experiments}

We compute the results for each detailed objective using the expressions shown in Table~\ref{table:computation} and compare it, where possible, with independently available ground truth. This helps validate both the modeling and analysis aspects of the approach proposed. We will also present results to highlight the efficiency of the decoupling approach. Structure- and semantics-preserving aspects of the decoupling approach allows us to drill down and show detailed experimental results.

\subsection{\textcolor{blue}{\em {US Airline  Analysis Results}}}
\label{sec:exp-results}
We now discuss the analysis results for the US airlines.


{\bf A1: Rank the top five cities for each carrier, that have the highest coverage.} For this analysis we computed the closeness centrality for each layer. We ranked the cities in each layer according to their closeness centrality value.

Top 5 hubs (\textit{higher rank, fewer flights required for coverage, more central city})) were identified for each airline.  {\em For all 6 airlines, the ground truth obtained from~\cite{hubs} matched our results.} In Table  \ref{table:A1} we have listed top 5 hubs for 4 airlines. As a byproduct, it is interesting to see common hubs (highlighted) between airlines which is also verified by the ground truth.




\begin{table}[h!t]
\renewcommand{\arraystretch}{1}
\centering
\vspace{-10pt}
\subfloat[]{

    \begin{tabular}{|c|}
        \hline
        \textbf{American} \\
        \hline
        \hline
        \textit{Dallas} \\
        \hline
        \textit{Chicago} \\
        \hline
        Charlotte \\
        \hline
        Philadelphia \\
        \hline
        Phoenix \\
        \hline
        \end{tabular}
        \label{table:A1-results-AA}
}
\subfloat[]{

    \begin{tabular}{|c|}
        \hline
        \textbf{Southwest} \\
        \hline
        \hline
        \textit{Chicago} \\
        \hline
        \textbf{Denver} \\
        \hline
        Baltimore \\
        \hline
        \textit{Dallas}  \\
        \hline
        \texttt{Las Vegas}  \\
        \hline
        \end{tabular}
        \label{table:A1-results-SW}
}
\subfloat[]{

    \begin{tabular}{|c|}
        \hline
        \textbf{Frontier} \\
        \hline
        \hline
        \textbf{Denver} \\
        \hline
         \texttt{Orlando} \\
        \hline
         Austin \\
        \hline
         \texttt{\textbf{Las Vegas}} \\
        \hline
         Philadelphia \\
        \hline
        \end{tabular}
        \label{table:A1-results-FT}
}
\subfloat[]{

    \begin{tabular}{|c|}
        \hline
        \textbf{Spirit} \\
        \hline
        \hline
        Fort Lauderdale \\
        \hline
         \texttt{\textbf{Las Vegas}} \\
        \hline
         \texttt{Orlando} \\
        \hline
         Detroit \\
        \hline
         \textit{Chicago} \\
        \hline
        \end{tabular}
        \label{table:A1-results-SP}
}

\caption{{\bf(A1)}Cities With Maximum US Travel Coverage}
\label{table:A1}
\vspace{-10pt}
    
\end{table}

{\bf A2: Classify airlines into major and minor carriers.}  There are several ways that this classification can be done. By simply looking at the edges in each layer, we can see the number of flights -- larger carriers will have more edges, hence more flights. A more pertinent classification is via computing the average degree. The more the average degree of a layer, more the connectivity of the corresponding airline.

The average degree for each airlines, given in the parenthesis, is as follows;  {\em American} (0.2622),  {\em Southwest} (0.04995),  {\em Delta} (0.2552),  {\em Frontier} (0.0384), {\em Spirit} (0.009995) and {\em Allegiant} (0.0701). Ordering each airline (layer) by their average degree shows a clear division of  \textit{American, Southwest and Delta} as {\em Major Airlines}; \textit{Allegiant, Frontier and Spirit} as {\em Minor Airlines}. This division also holds when we consider only the edges in each layer. 

This classification can be easily validated  using orthogonal data on {fleet size, revenue and passengers carried in a year} from~\cite{largest-airlines}.
We also identified the {common important operating bases}, that is, cities with {higher than average degree criteria} for both major and minor airlines. We found that, in general most of {\em such cities for minor airlines (Tampa, Orlando, Fort Lauderdale, Cleveland, ...) are smaller cities} in terms of {population and GDP per capita} as compared to the hubs of major airline (Dallas, Chicago, Los Angeles, New York, ...). Thus our results show that major and minor airlines focus on {different types of regions and demographics within the US}.

{\bf A3: Predict next hub for a carrier.} We chose Allegiant as the target minor airline which is considering expansion. The remaining airlines are chosen as competitors. 
Intuitively, the {cities must be considered for expansion} that are a) not yet a hub, b) have high coverage, i.e. high values of closeness centrality, this helps to reduce cost of expansion and c) do not have large operations, i.e. low closeness centrality, for the competitor airlines, this helps to minimize competition. 
From the high closeness centrality cities of the target airline, we removed the actual hubs first, followed by all those cities that are also high closeness in each of the competitor airlines. We then ranked this set of cities based on their high population.


\begin{wraptable}{l}{0.3\linewidth}
\small
\centering
    \begin{tabular}{|c|}
        \hline
        \textbf{Allegiant} \textit{Vs. All}\\
        \hline
        \hline
        \textbf{Grand Rapids} \\
        \hline
        Elko \\
        \hline
        Montrose \\
        \hline
        \end{tabular}
        \label{table:A3-results-AGvALL}
\caption{{\bf A3}: List of cities where Allegiant is likely to create its next hub.} 
\label{table:A3}
\vspace{-10pt}
\end{wraptable}

Table \ref{table:A3} shows the resulting set of cities where Allegiant Airline can potentially expand its operations. 
We validate our result by the fact that \textit{\textbf{Grand Rapids has been converted to a hub by Allegiant as of July 6, 2019}} \cite{AG-hub}. 

\subsection{\textcolor{cyan}{\em{IMDb HoMLN Analysis Results}}} To create this dataset we selected  the top 500 actors, we then extracted the movies they have worked in (7500+ movies with 4500+ directors). The actor set was repopulated with the co-actors from these movies, giving a total of 9000+ actors. 
We used the Louvain method (\cite{DBLP:Louvain}) to detect the layer-wise communities (partial results.) 


Around 44\% actors (mostly world renowned) had an average movie rating in the range [6-7) making it the \textit{most popular IMDb rating class}, while only 1.8\% actors have the highest average rating in the range [9-10]. On the other hand, the largest co-acting and similar genre groups had 15.6\% and 15.3\% actors, respectively.

{\bf A4: Find co-actors who have acted in similarly rated movies.} 2430 actor groups with similar average ratings were detected in which \textit{most of the actor pairs} have worked with each other. Few observations on the results:
\begin{itemize}
    \item For the most popular average actor rating, [6-7), the largest co-actor groups were from Hollywood (876 actors), Indian (44 actors), Hong Kong (12 actors) and Spanish (9 actors) movies.
    \item Among the Hollywood movie based groups, the top group included co-actors  {Al Pacino, Robert De Niro, and Will Smith}. Pacino and  De Niro acted together in famous movies like Heat and Godfather Part II. Interestingly, co-actors from less known movies, such as Smith and De Niro, in Shark Tale also come up.
    \item Famous Bollywood stars like {Amitabh Bachchan}, {Shah Rukh Khan} belonged to largest top rated Indian group. They acted together in many highly rated bollywood movies. 
    \item {Jackie Chan} (along with other lesser known actors) was among the prominent actors from the co-actor group from Hong Kong.
\end{itemize}


{\bf A5: Find actors who have worked in the same genre, but have not acted together.} We detected 900 groups of actors with similar genre preferences and average rating {but most of whom have not worked together}.  Table \ref{table:actorcollab} shows a few recognizable \textit{actors}  {who have not acted together}. Out of these,  as per reports in 2017,  there had been {\bf talks of casting Johnny Depp and Tom Cruise in pivotal roles in Universal Studios' cinematic universe titled Dark Universe} \cite{dark-universe}.

\begin{table}[h!t]
\renewcommand{\arraystretch}{1}
\centering
    \begin{tabular}{|p{5.4cm}|p{2.5cm}|}
        \hline
        \textbf{Actors/Actresses} & \textbf{Common Prominent Genres} \\
        \hline
        Willem Dafoe, Russell Crowe & Action, Crime\\
        \hline
        Hilary Swank, Kate Winslet & Drama \\
        \hline
        Tom Hanks, Reese Witherspoon, Cameron Diaz & Comedy, Romance\\
        \hline
        \textcolor{blue}{Johnny Depp, Tom Cruise} & \textcolor{blue}{Adventure, Action}\\
        \hline
        Leonardo DiCaprio, Ryan Gosling & Crime, Romance\\
        \hline
        Nicolas Cage, Antonio Banderas & Action, Thriller \\
        \hline
        Hugh Grant, Kate Hudson, Emma Stone & Comedy, Romance \\

        \hline
        \end{tabular}
        
\caption{{\bf (A6)}:  Highly rated genre actors who have not co-acted}
\label{table:actorcollab}
\end{table}

\subsection{\textcolor{violet}{{IMDb HeMLN Analysis Results} }}

For the same set of 7500+ movies used above, a HeMLN was built with an actor layer (9000+ actors), director layer (4500+ directors) and a movie layer. 
Louvain algorithm generated 63 Actor (A) and 61 Director (D) communities based on similar genres. Out of the 10 ranges (communities) in the movie (M) layer, most of the movies received a rating in the range [6-7), while least popular rating was [1-2). 

{\bf A6: Find groups of actors and directors who collaborate together.} 49 A-D (Actor-Director) similar genre-based community pairs are obtained, where {most actor-director pairs have interacted with each other at least once}. Intuitively, a group of actors that prominently works in some genre (say, Drama, Action, Romance, ...) must pair up with the group of directors who primarily make movies in the \textit{same genre}.  

\begin{figure*}[ht]
   \centering \includegraphics[width=\linewidth]{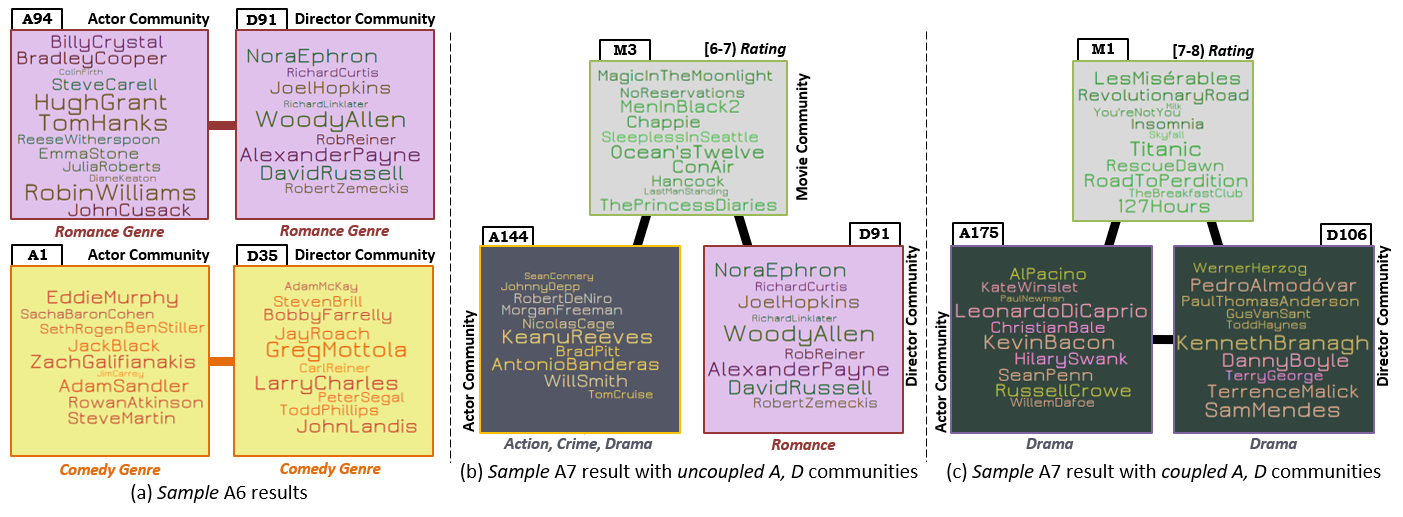}
  \vspace{-20pt}
   \caption{\textit{Sample} community match results for IMDb AMD HeMLN Analysis}
   \label{fig:imdb-hemln}
\end{figure*}

In Fig.\ref{fig:imdb-hemln} (a) we have shown A-D community pairings for the Romance and Comedy genres. Few famous actors and directors from each community have been listed. Such pairings may help production houses to sign up actors and directors for different movie genres. Recently, \textbf{Vin Diesel signed up for Avatar 2 and 3 (Action movie) which is being directed by James Cameroon and this will be the first time they will be collaborating}~\cite{avatar2}. Interestingly, even though they did not work together ever, we paired  them together in the groups that corresponded to the Action genre on the basis of {high interaction among other similar actors and directors}~\footnote{This pairing is not shown in the Fig.\ref{fig:imdb-hemln} due to space constraints}. 


{\bf A7: Find groups of actors and directors who collaborate together in highly rated movies.} 
When finding the communities across three layers, using the expression in Table~\ref{table:computation}, we first tried to combine results each of two layers with that of a common layer. Figure~\ref{fig:imdb-hemln} (b) shows the  results of one such combination, where actors (community A144) and directors (community D91) is paired with movies (community M3). However, we see that most popular actor and director groups for [6-7) movie rating (represented by M3) do not have many interactions. Even though few actor-director pairs from these two have collaborated on a few movies, it can be seen from Figure \ref{fig:imdb-hemln} (a) that D91 (community id in layer D) pairs (has maximum interaction) with A94, thus validating the absence of pairing between D91 and A144 in this result.

This result motivated us to couple the results of the three layers. Here the interactions between actors and movie communities are calculated, then interactions between movie and director communities are calculated and finally interactions between directors and actors are calculated. Only the communities for which the coupled paths (actor-movie-director-actor) are obtained are retained as results.



Only one HeMLN community shown in Figure~\ref{fig:imdb-hemln}(c) was obtained, which was not an extension of the previous result. This shows the capability of the HeMLN community detection to identify communities that \textbf{strongly interact cyclically in all 3 layers.} As the drill-down of Figure \ref{fig:imdb-hemln} (c) indicates, {both the popular groups for [7-8) movie rating are from Drama genre and many of these actor-director pairs have collaborated on many movies}, such as {\bf Leonardo DiCaprio, Kate Winslet with Sam Mendes for Revolutionary Road, Sean Penn with Gus Van Sant for Milk} and so on. Thus, the popular groups A175 and D106 have been  paired with each other.



\subsection{Efficiency Analysis of the Decoupling Approach}
\label{sec:exp-efficiency}

\noindent\textbf{Experimental Set up.} We used a quad-core 8th generation Intel i7 processor Linux machine with 8 GB memory. 
The layer-wise results (communities or hubs) are generated \textit{once} and can be done in \textit{parallel}. Thus, this one time cost is bounded by the layer that takes maximum time. Moreover, the cost of composing the partial results using Boolean AND (HoMLN Hubs and Communities) or Maximum Weighted Bipartite Matching (HeMLN communities) is \textit{significantly less} than recomputing over the combined MLN layers. 

For HoMLN analysis, we compare the total computational cost of the decoupling approach and the traditional single graph approach which includes the {time to generate the combined layer} followed by generating the degree/closeness hubs or communities. For HeMLN, we show how the \textit{incremental cost} of generating the community pairings is minimal.

\begin{figure}[h]
\centering
\includegraphics[width=\linewidth]{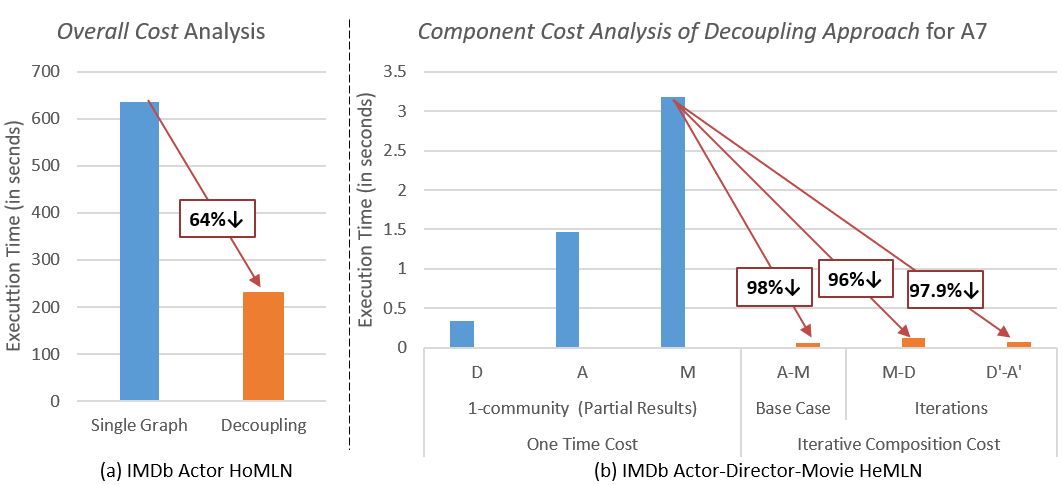}
\caption{Efficiency of Decoupling Approach for MLN analysis}
	\label{fig:MLN-eff}
\end{figure}


\textcolor{cyan}{\textit{IMDb HoMLN Analysis Cost}}: Two Boolean AND compositions are required as per Table \ref{table:computation} to generate communities. Here the one time cost for finding the layer-wise communities is bound by the \textit{AvgRating} layer (densest layer in Table \ref{table:IMDbHoMLNStats}).
Overall, \textbf{64\% reduction in computation time} is observed with the \textit{decoupling approach (228.865 seconds)} as compared to \textit{single graph approach (636.45 seconds)} as in Figure \ref{fig:MLN-eff} (a).

US Airline analysis which is also a HoMLN ( hub composition instead of community composition) showed a \textbf{reduction of 33.6\% in computation time}. 

\textcolor{violet}{\textbf{IMDb ADM Analysis Cost}}: Figure \ref{fig:MLN-eff} (b) shows the execution time for the \textit{one-time} cost of analyzing each layer, and \textit{iterative composition costs} for A7 - the most complex HeMLN community we have computed which uses 3 iterations of pairings. The difference in one-time cost for the 3 layers matches their edge density, as given in Table \ref{table:IMDbHeMLNStats}. Iteration cost includes {\em (i)} creating the bipartite graph with the layer-communities that are part of the result , {\em (ii)} computing meta edge weights, and {\em (iii)} cost of computing MWBM. The iterative cost is insignificant as compared to the one time cost (by an order of magnitude.) 
Even {the cost of all iterations together (0.2582 sec) is still almost {an order of magnitude less than the largest one-time cost} (3.173 sec for Movie layer.)} 
The {incremental cost for computing a community in HeMLNs is extremely small}. 

\section{Related Work}
\label{sec:related-work}

 We present related work for analyzing homogeneous and heterogeneous MLNs based on community and hub identification.

\textbf{Analysis of HoMLNs:} \textit{Community detection} algorithms have been extended to {HoMLNs} for identifying groups of tightly knit nodes based on different feature combinations (see reviews \cite{CommSurveyKimL15,CommFortunatoC09}). 
 Algorithms based on matrix factorization, 
cluster expansion,  
Bayesian probabilistic models,  
regression, 
and spectral optimization of the modularity function based on the supra-adjacency representation 
have been developed. Further, some methods have been developed to determine \textit{centrality measures} to identify highly influential entities as well \cite{de2013centrality,sole2014centrality,zhan2015influence}. However, all these approaches \textit{analyze a MLN  by reducing it to a simple graph} either by aggregating all (or a subset of) layers or by considering the entire multiplex as a whole, thus leading to loss of semantics as the entity and feature type information is lost.  

Recently, our group has proposed decoupling-based approaches for detecting communities~\cite{ICCS/SantraBC17} and centrality~\cite{ICDMW/SantraBC17} in HoMLN, where partial analysis results from individual layers are combined {systematically in a loss-less manner} to compute communities or centrality hubs for combinations of layers. Due to the "divide and conquer" approach of decoupling, this method  is more efficient as it avoids re-computation of layer communities and also provides flexibility of analysis. 


\textbf{Analysis of HeMLNs:} Majority of the work on analyzing HeMLN (reviewed in \cite{shi2017survey,sun2013mining}) focuses on developing meta-path based techniques for determining the 
classification of objects, 
predicting the missing links, 
ranking 
and recommendations
. A few existing works have proposed techniques for generating clusters of entities \cite{melamed2014community}. 
However,  most of these methods concentrate mainly on the inter-layer edges and not the networks themselves. Moreover, the existing approaches (type-independent~\cite{LayerAggDomenicoNAL14} and projection~\cite{Berenstein2016}) do not preserve the structure or types and labels of nodes and edges. The type independent approach collapses  all layers into a single graph keeping \textit{all} nodes and edges (including inter-layer edges) sans their types and labels. Similarly,  the projection-based approach projects the nodes of one layer onto another layer and uses the layer neighbor and inter-layer edges to collapse the two layers into a single graph with a single entity type instead of two.

In this paper, for the HoMLN community and hub detection, we use the algorithms  in~\cite{ICCS/SantraBC17,ICDMW/SantraBC17}. For HeMLN, we use a  {structure-preserving HeMLN community detection} that takes into account the {combined effect of layer communities, entity types, intra- and inter-layer relationships (types)} as in~\cite{Arxiv/SantraKBC19}. 

\section{Conclusions}
\label{sec:conclusions}

We have argued for using MLNs for modeling as well as structure- and semantics-preserving analysis using the decoupling approach. We believe that the diffidence in using MLNs comes from lack of composition algorithms as compared to other modeling alternatives. We have used recent work on efficient community and hub composition approaches for MLNs. We have applied it on two data sets to cover different types of complex data to demonstrate its versatility of analysis as well as computational efficiency using the decoupling approach.

In future we aim to extend this framework for other computations, such as subgraph mining and link prediction, on HoMLN and HeMLNs on more complex data sets, with disparate data types and greater number of layers.

\begin{scriptsize}
\bibliographystyle{plain}
\bibliography{./bibliography/somu_research,./bibliography/itlabPublications,./bibliography/itlabTheses,./bibliography/santraResearch,./bibliography/itlabPublications.original}
\end{scriptsize}

\end{document}